\documentclass[twocolumn,preprintnumbers,amsmath,amssymb]{revtex4}
\usepackage{dcolumn}
\usepackage{bm}
\usepackage{subfig}
\usepackage[pdftex]{graphicx}
\graphicspath{{images/}}

\begin{document}


\title{Equilibrium Geometries, Reaction Pathways, and Electronic Structures of Ethanol Adsorbed on the Si~(111) Surface}

\author{Alexander V. Gavrilenko}
\affiliation{Center for Materials Research, Norfolk State University, 700 Park Ave, Norfolk, VA 23504}
\author{Carl E. Bonner}
\affiliation{Center for Materials Research, Norfolk State University, 700 Park Ave, Norfolk, VA 23504}
\author{Vladimir I. Gavrilenko}
\affiliation{Center for Materials Research, Norfolk State University, 700 Park Ave, Norfolk, VA 23504}

\begin{abstract}
Equilibrium atomic configurations and electron energy structure of ethanol adsorbed on the Si~(111) surface are studied by the first-principles density functional theory. Geometry optimization is performed by the total energy minimization method. Several equilibrium atomic configurations of ethanol, both undissociated and dissociated, on the Si~(111) surface are found. Reaction pathways and predicted transition states are discussed in comparison with available experimental data in terms of the feasibility of the reactions occurring. Analysis of atom and orbital resolved projected density of states indicate substantial modifications of the Si surface valence and conduction bands due to the adsorption of ethanol affecting the electrical properties of the surface.
\end{abstract}


\maketitle

\section{\label{sec:intro}Introduction}
Physics of organic molecules adsorbed on semiconductor surfaces opens up a new field of applications in high-tech industry such as chemical sensors and molecular electronic devices \cite{wolkow99,bent02,filler03}. A detailed understanding of the adsorption mechanisms of hydro-carbons on solid crystalline surfaces is one of the most important topics of modern material and surface science (see e.g. \cite{bechstedt03} and references therein). State-of-the-art first principle methods based on the Density Functional Theory (DFT) are very useful tools for providing a detailed understanding of structural and electronic processes on solid surfaces \cite{bechstedt03,nunzi04,silvestrelli04}. 

Silicon is one of the basic materials used in modern electronics. The atomic processes on Si surfaces have attracted interest from both fundamental and applied aspects of the surface science \cite{bechstedt03}. Adsorption of organic molecules on Si surfaces has been extensively studied both experimentally \cite{bulanin01,dilabio06,eng97} and theoretically \cite{eng97,biering06,cho04,nemanick06}. Equilibrium atomic structures, thermodynamic stabilities, and transition states have been studied within last decade \cite{zhang03,miotto05,silvestrelli01,romero03,costanzo03,silvestrelli02,wieferink06}. Fourier Transform Infrared (FTIR) spectroscopy and Scanning Electron Microscopy (SEM) studies demonstrated dissociative character of organic molecular adsorption on Si-surface \cite{bulanin01,dilabio06}. Due to its applications in microelectronics a great deal of attention was given to the study adsorption mechanisms on the Si~(001) surface \cite{widdra96,miotto02,silvestrelli00,lee05,brown99}.

Ethanol is a relatively small organic molecule, which represents both a model system for the investigation of organic molecules with silicon surface, but which is also of fundamental importance since ethanol is used in many processing (cleaning) steps in the preparation of silicon surfaces for a variety of technical applications. Nonetheless, little theoretical work has been performed to underpin experimental studies of the interaction of ethanol with Si. Previous ethanol adsorption studies by Silvestrelli \cite{silvestrelli04}, Eng \cite{eng97}, and Zhang \cite{zhang03}  were done on the Si~(100) surface concluding that the dissociation kinetics are energetically favored by O-H bond cleavage above the Si-Si dimer. This adsorption mechanism cannot be applied to the Si~(111) surface because it is completely different from the Si~(001) surface with respect to the geometric and electronic structure. The Si dimers which influence and, to a large extent, define the adsorption mechanisms observed on the Si~(001) surface are not present on the Si~(111) surface. However, the charge separation that occurs on the Si-Si dimers, which is also responsible for the buckling observed, occurs to and extent on the Si~(111) surface. The Si-Si dimers, consisting of an \emph{up} nucleophillic and a \emph{down} electrophillic Si atoms, attract through Coulombic forces the ethanol molecule, which has significant charge separation of its own, forming a \emph{dative} bond between the \emph{down} Si atom and ethanol's oxygen. At this stage the molecule is providing both electrons required for the bond. In the presence of ethanol similar buckling and charge separation is observed on the Si~(111) surface, which unlike the Si~(100) surface does not have charge separation nor buckling in the absence or the organic molecule. Due to the Si surface atoms longing for a stable configuration filling their outer \emph{p}-shell and due to the potential exerted by the ethanol molecule, charge separation occurs on the Si~(111) surface which in turn causes buckling forming a similar configuration to the Si~(100) surface with a \emph{down} Si atom immediately below ethanol's O-H group and alternating in the four directions throughout the surface. This can only be observed on an unpassivated Si~(111) surface where the terminating hydrogens were desorbed previously.

Nanostructured materials, for example porous silicon (PorSi), have experienced a recent surge in interest due to their unique properties and applications. PorSi, consisting of a network of pores of various sizes embedded in silicon layer, was discovered to have strong photoluminesence (PL) at room temperature in the visible spectrum \cite{canham90} due to quantum confinement \cite{ohno92}. The luminescence from PorSi has been observed to be up to five orders of magnitude stronger than that of bulk Si \cite{fauchet93}. The enhanced PL emission intensity of PorSi is in part caused by surface chemistry \cite{munekuni90} which shows strong sensitivity to the environmental presence of various molecules leading to variations of the PL intensity and peak shift \cite{balarin07, baratto00, benilov07, zhao05}, impedance \cite{galeazzo03}, capacitance \cite{li07}, I-V characteristics \cite{irajizad04}, and Fabry-Perot fringe patterns \cite{min00, stefano03}. The large surface to volume ratio of PorSi causes the changes on the surface to affect the properties of the material as a whole. Thus detailed study of adsorption mechanisms on the surface is even more crucial for nanostructured materials.

The work presented here is organized as follows: section \ref{sec:method} describes the computational method. Section \ref{sec:results} presents the result separated into three subsections. The predicted equilibrium atomic configurations are described in section \ref{subsec:configs}. Transition states along with reaction coordinate graphs for most favorable reaction paths are presented and discussed in section \ref{subsec:transstates}. In section \ref{subsec:elecstruc} the spectra of atom and orbital projected density of electronic states are analyzed.

\section{\label{sec:method}Method}
Density functional theory (DFT) has been shown for decades to be very successful in the ground state analysis of different materials \cite{kohn96}. For atomic system which include molecules and solids it is important to realistically describe both short (covalent) and long range (Coulomb and van der Waals, vdW) components of intermolecular and molecule-surface interactions \cite{ortmann05}. Local density (LDA) \cite{perdew92} and generalized gradient approximations (GGA) \cite{burke96} are frequently used methods to account for exchange and correlation (XC) interaction. The vdW interactions is not included in standard DFT. However detailed analysis of organic molecule adsorption on solid surface demonstrates that around equilibrium the kinetic energy of valence electrons remains mainly repulsive, and XC effects are mostly responsible for the attraction \cite{ortmann05}. It has been shown that equilibrium distance between organic molecule and solid (graphene) surface predicted by LDA is in good agreement to the value followed from explicit inclusion of the vdW into interaction Hamiltonian \cite{ortmann05}. 

In this work we used \emph{ab initio} pseudopotential method within the DFT and the supercell formalism to study equilibrium geometry, energy kinetics parameters of dissociative chemical reactions, and electron energy structure of Ethanol adsorbed on Si~(111) surface. The unreconstructed Si~(111) surface is unstable which results in its further reconstruction \cite{bechstedt03}. This surface is also rather reactive and hydrogen adsorption is frequently used to passivate it. Our adsorption study of ethanol on such Si~(111) surfaces could be separated into two phases: first reactivation of the surface and then the reaction of ethanol with dangling bonds. Here we concentrate on the second phase and consider unreconstructed Si~(111) surface. The equilibrium geometries of CH$_{3}$CH$_{2}$OH molecule adsorbed on the Si~(111) unreconstructed surface are calculated by LDA method \cite{perdew92} employing norm-conserving pseudopotentials \cite{hamann79} using the DMol$^{3}$ \cite{delley00} software package. We have explicitly checked that the structural and binding properties of our system are well converged for the double numerical plus polarization basis set used. The use of the exact DFT spherical atomic orbitals has several advantages. For one, the molecule can be dissociated exactly to its constituent atoms (within the DFT context). Because of the local character of these orbitals, basis set superposition effects \cite{delley90} are minimized and good accuracy is achieved even for weak bonds. Absolute binding energy values of the equilibrium structures are computed within GGA employing ultrasoft pseudopotentials \cite{vanderbilt90} and by using the CASTEP \cite{segall02} software package. The transition states, energy barriers, and reaction coordinates, have been obtained using the ``generalized synchronous transit method'' \cite{govind03}.

\begin{figure}[t]
\centering
\includegraphics[scale=0.15]{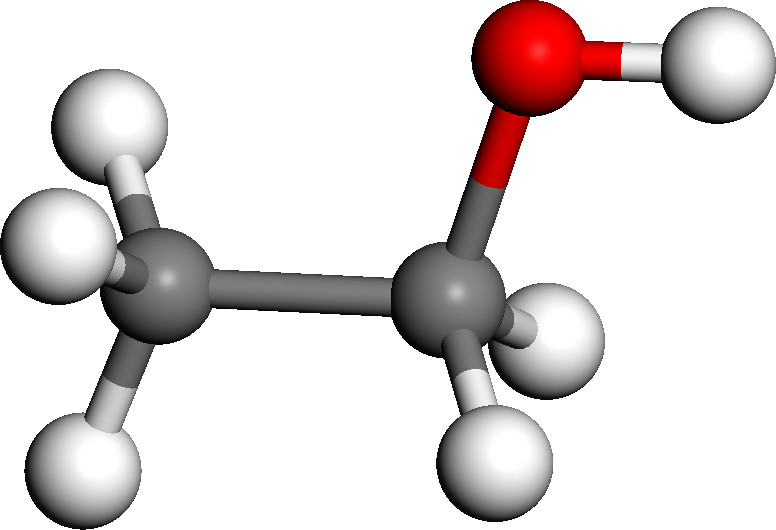}\\
\caption{\label{fig:isolated-ethanol-img}(Color) Isolated ethanol molecule. The white, red, and grey, spheres represent hydrogen, oxygen, and carbon atoms, respectively.}
\end{figure}

For free-standing ethanol molecule the method chosen in this work shows excellent agreement with experimental results. The structure of the isolated molecule is shown in Fig.~\ref{fig:isolated-ethanol-img}, while geometrical parameters are reported in Table~\ref{tab:isolated-ethanol-tbl}.

\begin{table}[t]
\caption{Predicted structural parameter of the free ethanol molecule in comparison with experimental data (see Fig.~\ref{fig:isolated-ethanol-img})}
\label{tab:isolated-ethanol-tbl}
\begin{ruledtabular}
\begin{tabular}{ l c c }
{}&Calculations&Experiment \cite{lide00}\\\hline 
\emph{d}(O-H) (\AA)&0.97&0.97\\
\emph{d}(C-O) (\AA)&1.41&1.43\\
\emph{d}(C-C) (\AA)&1.50&1.51\\
$\angle$C-C-O (deg)&107.9&107.8\\
$\angle$C-O-H (deg)&108.5&105.0\\
$\angle$C-C-H (deg)&109.9&110.0\\
$\angle$C-C-H (deg)&110.9&111.0\\ \hline
\multicolumn{3}{c}{C-H bonds have lengths in the range 1.101-1.11.}\\ 
\end{tabular}\\
\end{ruledtabular}
\end{table}

The solid surface is modeled as a $(2\times2)$ periodic structure of six Si bi-layers and a 12~{\AA} thick vacuum layer separating the slabs to eliminate the spurious interactions between the ethanol molecule and back face of the slab; the bottom four Si bi-layers are constrained in order to reproduce the bulk electronic structure of Si. For the surface physics contribution of intermolecular interaction on the surface is important. However, by chosing the $(2\times2)$ unit cell (where adsorbed molecules are far apart from each other) this issue remains out of scope in this paper. Hydrogen atoms are used to saturate the dangling bonds on the back surface of the slab which are also allowed to relad providing a measure of interaction with the ethanol molecule from the unitcell below. The adsorption energy per molecule is defined as the difference between the total energy of the adsorption system and the energies of the isolated components, namely the clean substrate and the adsorbate, divided by the number of adsorbed molecules \emph{N} per unit cell:

\begin{eqnarray}
E_{ads}=(E_{substrate}+E_{adsorbate}-E_{tot})/N
\end{eqnarray}
Note that the adsorption energy is positive by definition when the adsorbates bind to the substrate.

\section{\label{sec:results}Results and Discussion}
In this section we present first the equilibrium geometries resulted by atomic relaxation during the adsorption of ethanol on Si~(111) surface. The second part of this section discusses the predicted transition states while the related transformations of electronic structure are discussed in the third part of this section.

\subsection{\label{subsec:configs}Equilibrium Configurations}

\begin{figure*}[ht]
\centering
\includegraphics[width=0.8\textwidth]{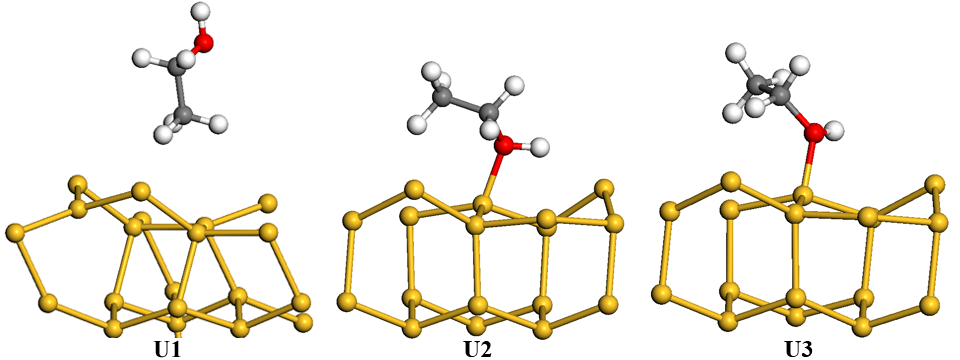}\\
\caption{(Color) Stable configurations for physisorption of ethanol on Si~(111). As before, the white, red, and grey, spheres represent hydrogen, oxygen, and carbon atoms, respectively, while the yellow spheres represent silicon atoms.}
\label{fig:ethanol-physisorbed-img}
\par
\end{figure*}

\begin{table*}[ht]
\caption{Binding energies and structural parameters of the configurations shown in Figs.~\ref{fig:ethanol-physisorbed-img}
and \ref{fig:ethanol-chemisorbed-img}}
\label{tab:energies-structures-tbl}
\begin{ruledtabular}
\begin{tabular}{ @{}l|c c c|c c c c c c|c c c}
{}&\multicolumn{3}{c|}{\underline{Undissociated}}&\multicolumn{6}{c|}{\underline{1 Step Dissociated}}&\multicolumn{3}{c}{\underline{2 Step Dissociated}}\\
{}&U1&U2&U3&D1&D2&D3&D4&D5&D6&D7&D8&D9\\\hline
$E_{ads}$ (eV)&0.93&1.524&1.522&3.25&3.26&2.33&4.00&2.46&2.59&3.57&4.29&4.25\\
\emph{d}(O-H) (\AA)&0.97&1.05&1.04&-&-&0.97&0.98&0.97&0.99&-&-&-\\
\emph{d}(C-O) (\AA)&1.41&1.46&1.46&1.42&1.42&1.41&-&1.41&1.40&1.41&1.44&1.45\\
\emph{d}(C-C) (\AA)&1.50&1.49&1.49&1.50&1.50&1.50&1.51&-&-&-&-&-\\
\emph{d}(C-Si) (\AA)&-&-&-&-&-&1.88&1.89&1.90&1.93&1.88&1.98&1.97\\
\emph{d}(O-Si) (\AA)&-&1.85&1.85&1.66&1.66&-&1.67&-&-&1.67&1.69&1.70\\
\emph{d}(H-Si) (\AA)&-&-&-&1.50&1.51&1.50&-&-&-&-&1.51&1.51\\
\end{tabular}
\end{ruledtabular}
\end{table*}

Predicted adsorption energies of Ethanol adsorbed on Si~(111) are presented in Table~\ref{tab:energies-structures-tbl}. Undissociated and dissociated atomic configurations are denoted with U and D, respectively. Physisorption of Ethanol with CH$_{3}$ group oriented downward to the surface is unfavorable (see U1 Fig.~\ref{fig:ethanol-physisorbed-img} and Table~\ref{tab:energies-structures-tbl}). Consequently the ethanol molecule favors the interaction with the Si~(111) surface through the formation of a dative bond between the O atom and a Si surface atom. In these states the ethanol molecule remains undissociated with different stable orientations with respect to the Si surface (see U2 and U3 in Fig.~\ref{fig:ethanol-physisorbed-img}).  Creation of dative bond results in geometry changes (as shown in Fig.~\ref{fig:ethanol-physisorbed-img}), charge redistribution between the O and Si surface atoms, and lowering of the total energy (see Table~\ref{tab:energies-structures-tbl}). This reaction does not have an energy barrier and occurs spontaneously at room temperature. The C-C/C-H bond cleavage has an energy barrier much higher than that related to the molecule rotation above the Si~(111) surface; the ethanol molecule will simply rotate into a more favorable, the 'OH group down', configuration.

\begin{figure*}[t]
\centering
\includegraphics[width=0.8\textwidth]{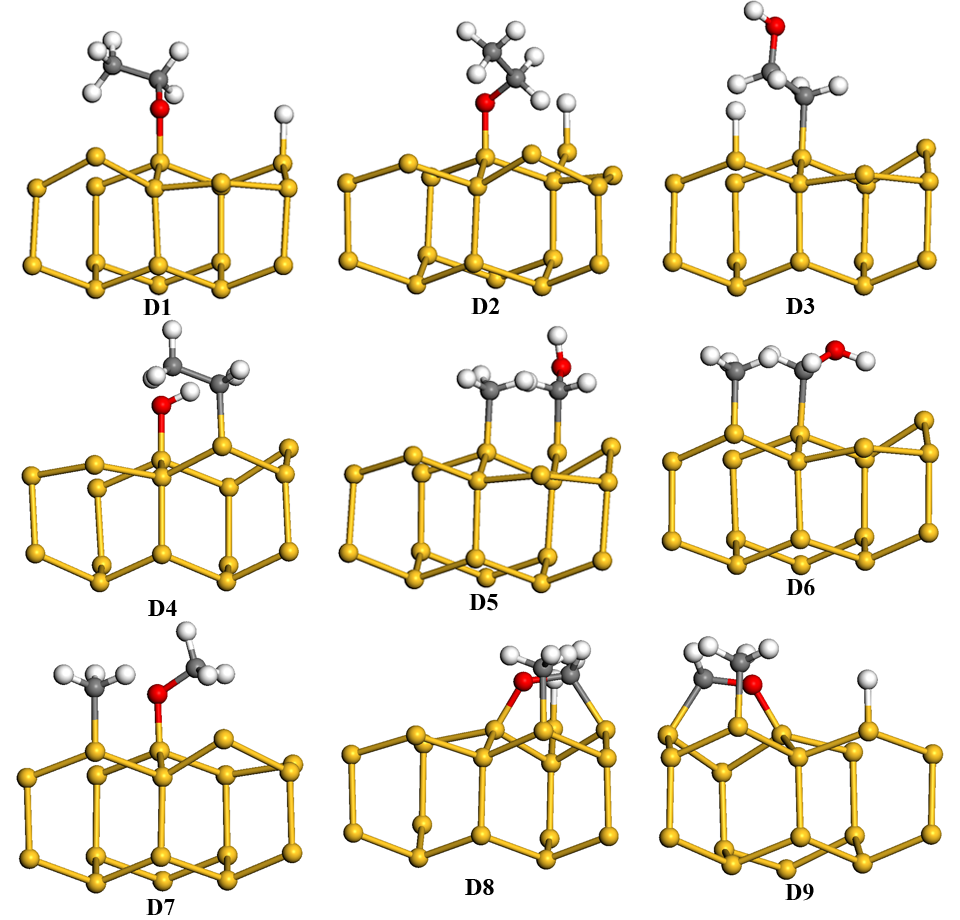}\\
\caption{(Color) Stable configurations for dissociative chemisorption of ethanol on Si~(111). As before, the white, red, grey, and yellow spheres represent hydrogen, oxygen, carbon, and silicon atoms, respectively.}
\label{fig:ethanol-chemisorbed-img}
\end{figure*}

The atomic configurations most energetically favorable appeared after ethanol dissociation on the surface (see Table~\ref{tab:energies-structures-tbl}). It is instructive to consider the energy of transition states on the path from undissociated to dissociated configurations. Starting from U1, U2, or U3 states, there is a number of possible pathways for continuous ethanol reactions with the Si surface, which imply breaking one or more molecule bonds in order to realize more stable, dissociative, configurations. We found nine stable dissociative configurations which are shown in Fig.~\ref{fig:ethanol-chemisorbed-img}. 

The D1 and D2 structures are obtained through breaking of the O-H bond while the detached proton forms a new bond with a Si surface atom. Both of these reactions have very small energy barriers and are ``self-sustaining'' by drawing energy from the initial adsorption to U2/U3. This will be discussed in more details in a subsequent paragraph. The D3 configuration is achieved through the C-H bond cleavage and forming bonds between Si-C and Si-H. The starting point for this reaction is the U1 configuration which is energetically significantly less favorable than U2/U3. In addition, D3 itself is energetically less favorable than the previously mentioned D1 and D2 configurations. Thus even though the D3 configuration is possible, it is much more likely to find the D1 and D2 configurations. The fourth dissociative configuration, D4, can be obtained by breaking the covalent C-O bond; the ethyl group attaches to another Si surface atom. This is the energetically most favorable configuration out of all the found dissociated structures which have two bonds between surface and molecule. The D5 and D6 structures are obtained through cleavage of the C-C bond, followed by the methyl and CH$_{2}$OH groups forming new bonds with two neighboring Si surface atoms. The starting point for this reaction is U1. Considering the energy required to break the C-C bond and the energy of the U1 and D5/D6 structures as compared to other configurations, it is highly unlikely that this configuration will be seen. The D7 structure is obtained through the C-C bond cleavage, just like D5 and D6, however this configuration follows from the dative undissociative configurations U2 and U3. The methyl group bonds to a Si surface atom, while the proton migrates from the oxygen to the danging bond on the carbon. This also converts the Si-O bond to a covalent bond. The two ``bridging'' structures D8 and D9 follow from the D2 and D3 configurations, cleaving the O-H bond; two new bonds are created between the oxygen and another Si surface atom and between the detached proton and yet another Si surface atom.  

\subsection{\label{subsec:transstates}Transition States}
\begin{figure*}[t]
\centering
\includegraphics[width=\textwidth]{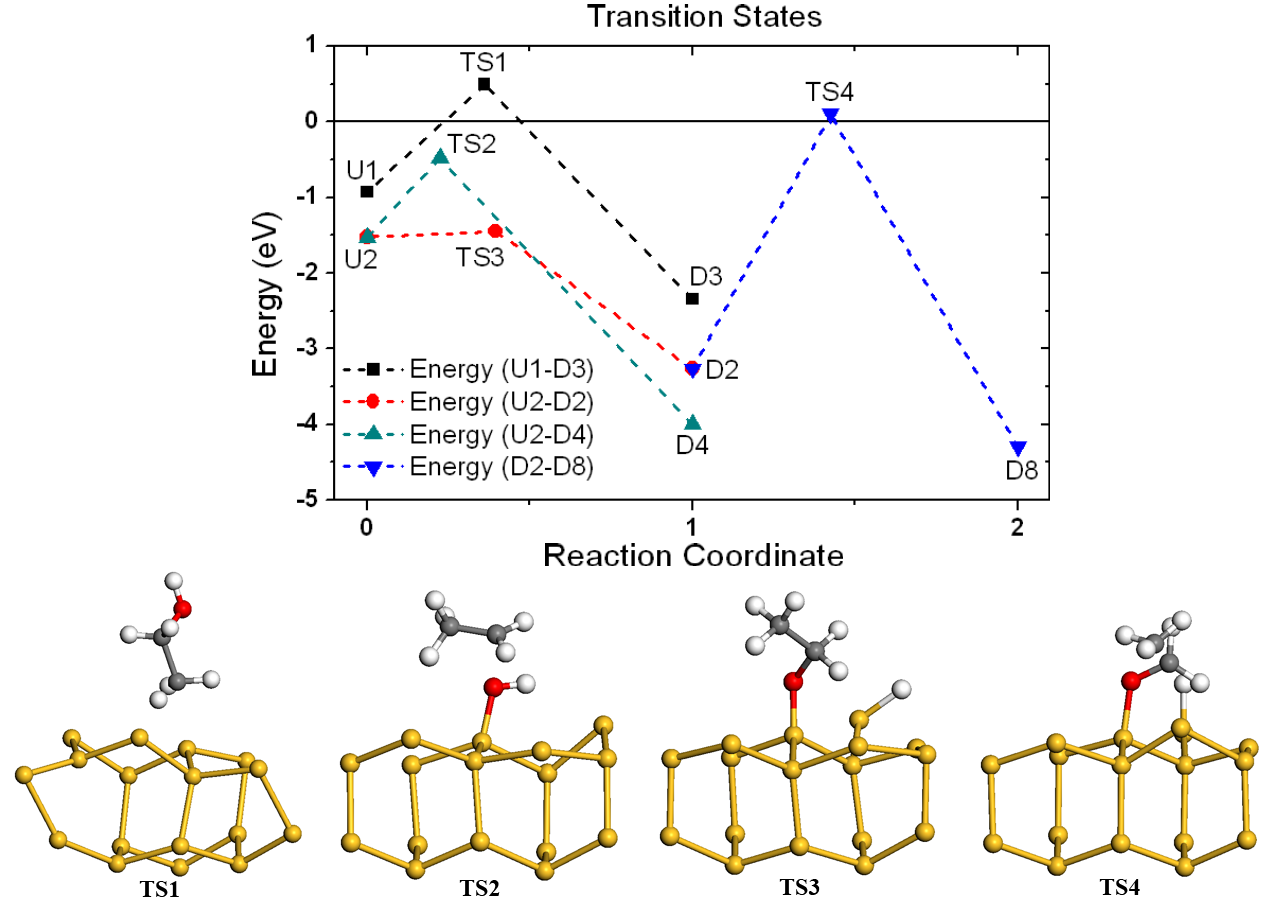}\\
\caption{(Color) Reaction coordinates and transition states for typical configurations. As before, the white, red, grey, and yellow spheres represent hydrogen, oxygen, carbon, and silicon atoms, respectively.}
\label{fig:tssearch-grph}
\end{figure*}

The values of activation energy barriers are important energy parameters for the surface reaction kinetics. In Fig.~\ref{fig:tssearch-grph}, the transition energy paths are given for a few types of the surface reactions. The transition state TS1 of the reaction U1-D3 is predicted to be above free-energy level, the energy of the system where the molecule is infinitely separated from the surface, which indicates that this kind of reaction leads to the desorption and it is less probable on the Si(111) surface.

Note that significant Si host atom reconstruction occurs in the U1 structure. Instead of the Si surface having bulk-like six-fold rings, the surface has alternating  of seven- and five-fold rings (see U1 in Fig.~\ref{fig:ethanol-physisorbed-img}). Under excitation the molecule undergoes a reaction and it arrives at D3 product configuration. The D3 configuration is more stable than U1 by about 1.4eV. However the energy barrier of over 1.42 eV is significant and will not occur during normal circumstances. 

The activation energy required to go from U2 to D4 is only 1eV (see TS2 in Fig.~\ref{fig:tssearch-grph}), which is significantly less than that required to go for U1 to D3 and it is below the free-standing energy level and thus can occur at room temperature. In addition, the starting structure U2 is energetically much more favorable than U1. The resulting configuration D4 is one of the most stable and energetically most favorable structures we found. 

The undissociated configuration U2 is transformed through dissociation of the Ethanol molecule into D2 configuration over a very small activation barrier. The energy reduction of the transition from U2 to D2 is 1.74 eV (see Fig.~\ref{fig:tssearch-grph}). The predicted energy of the transition state TS2 between U2 and D2 is 0.07 eV above U2. This indicates that dissociation of hydrogen and its reaction with the Si dangling bond is the most probable process at room temperature.  The molecule shown in D2 can then proceed to dissociate further into D8. Though the resulting structure is \emph{the} most stable structure we found the very high energy of the D2$\rightarrow$D8 transition step (3.36eV) would make this step very unlikely, especially since the activation energy for the backreaction D2$\rightarrow$U2 is smaller. Therefore it would seem likely that dissipation of the energy released in the U2$\rightarrow$D2 step will cause D2 to become a trap state and that even though D8 may have the lowest energy state, it will not be reached without \emph{significant} thermal or photo excitation. 

In order to understand the nature of surface electronic structure modifications through ethanol adsorption it is important to analyze the projected density of states (PDOS). Details of the PDOS spectra calculations are given in \cite{zhu07}. Electron energy structure changes during the U2-D2 reaction the \emph{s}- and \emph{p}-orbital PDOS spectra are given in Fig.~\ref{fig:ARPDOS} \protect\subref{subfig:s-totd2u2grph} and \protect\subref{subfig:p-totd2u2grph} for both the U2 and D2 configurations, respectively.

\subsection{\label{subsec:elecstruc}Electronic Structure}

\begin{figure*}[ht]
\centering
\subfloat[U2 and D2 configurations with individual atoms labeled.]{\includegraphics[scale=0.4]{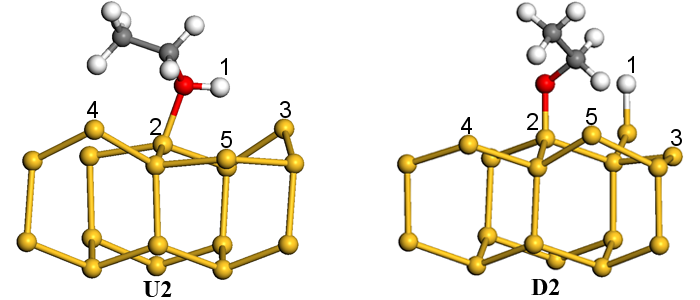}\label{subfig:u2d2labeled}}\\
\subfloat[\emph{S}-shell]{\includegraphics[width=0.45\textwidth]{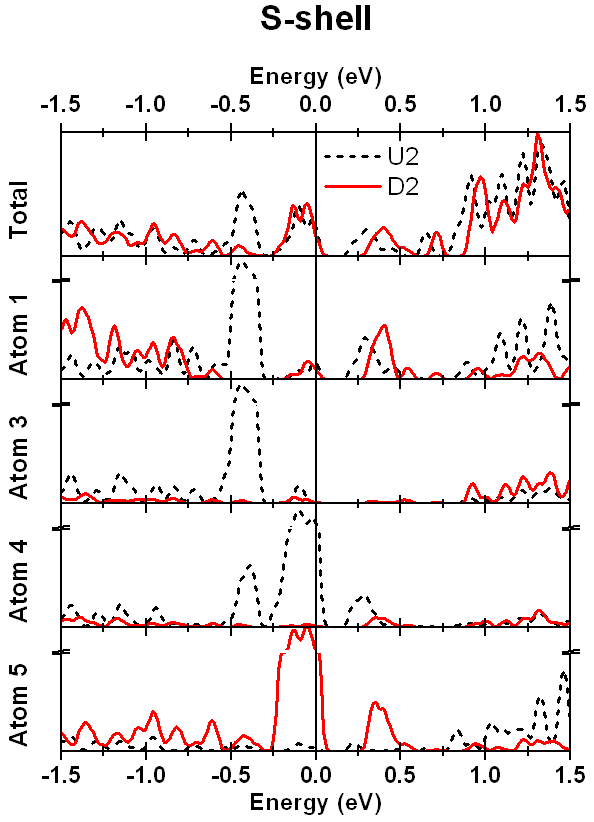}\label{subfig:s-totd2u2grph}}
\hspace{0.1in}
\subfloat[\emph{P}-shell]{\includegraphics[width=0.45\textwidth]{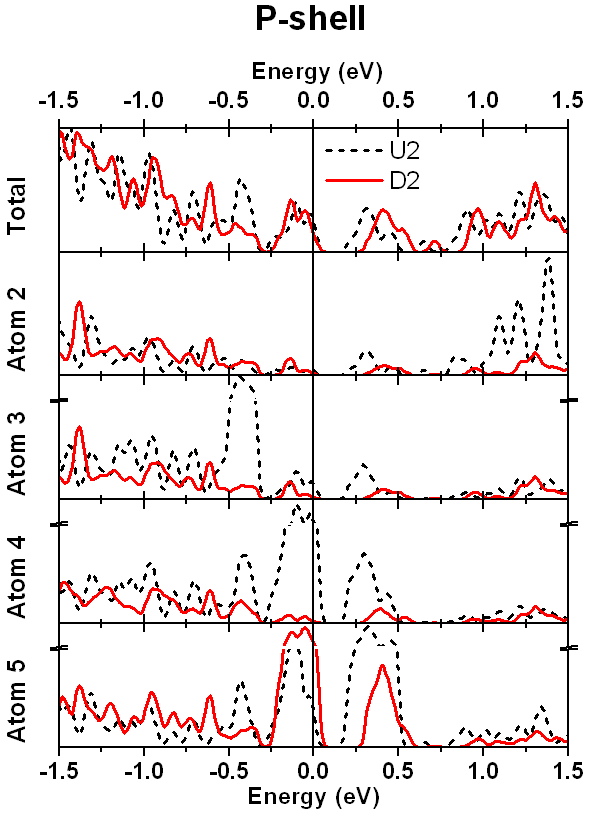}\label{subfig:p-totd2u2grph}}\\
\caption{(Color) Atom and shell resolved projected density of states for U2 and D2 configurations. Numbers on atomic geometry pictures \protect\subref{subfig:u2d2labeled} denote surface atoms which correspond to the panels of \emph{s}- \protect\subref{subfig:s-totd2u2grph} and \emph{p}-shell \protect\subref{subfig:p-totd2u2grph} PDOS spectra. As before, the white, red, grey, and yellow spheres represent hydrogen, oxygen, carbon, and silicon atoms, respectively. The dashed black and solid red lines (figures \protect\subref{subfig:s-totd2u2grph} and \protect\subref{subfig:p-totd2u2grph}) represent the U2 and D2 configurations, respectively.}
\label{fig:ARPDOS}
\end{figure*}

For bulk Si, our GGA-based computational method underestimates the gap by about 0.5 eV. A comparison of calculated PDOS spectra of U2 and D2 structures with those reported earlier for bare and hydrogenated Si~(111) surface \cite{bechstedt03} indicates that electronic states in the gap region around 0.2 to 0.5 eV correspond to unsaturated dangling bonds of surface Si atoms. These states are completely eliminated after saturation with hydrogen of the remaining Si dangling bonds on the surface. As expected the saturation of Si-dangling bonds with ethanolic hydrogen due to U2-D2 reaction (see Figs.~\ref{fig:ethanol-physisorbed-img} and \ref{fig:ethanol-chemisorbed-img}) results in change to the surface energy structure in the gap. Electron states changes in the bandgap (see \emph{total} panel in Fig.~\ref{fig:ARPDOS} \protect\subref{subfig:s-totd2u2grph} and \protect\subref{subfig:p-totd2u2grph}) appear due to hybridization caused by unpaired Si electrons on the surface (dangling bonds). 

Further, it is important to understand the hybridization changes of electronic bonds on the surface in the course of the surface reaction. Atom-resolved PDOS spectra provide with additional information about single unit cell atom contributions to the total density of states. In Fig.~\ref{fig:ARPDOS} we present both atom, as well as \emph{s}- and \emph{p}-orbital resolved PDOS spectra related for the reaction U2-D2. The U2 to D2 reaction path is energetically more preferable then other reactions studied here (see Table~\ref{tab:energies-structures-tbl}). It is interesting to understand how electronic structure is modified by the transformation of the bond between the surface and the molecule from dative to covalent through dissociation. 

The shape of atom resolved projected density of states spectra (AR-PDOS) shown in Fig.~\ref{fig:ARPDOS} reveals that there are no electronic states around zero energy related to the two ethanol carbons, oxygen, nor any of the carbon-bonded hydrogens. The electronic structure corresponding to these states is located far away from Fermi energy and is not discussed here. Fig.~\ref{fig:ARPDOS} shows the AR-PDOS for the U2 (dashed black lines) and D2 (solid red lines); only atoms that have electron states around 0 eV are shown. A comparison of the upper panels in Fig.~\ref{fig:ARPDOS} indicates that due to the dissociation of the hydrogen atom and the creation of the Si-H bond, the hydrogen related electronic states are pushed out of the gap. Unsaturated dangling bond states still present in the gap area around 0.4 eV do not show rehybridization during the U2-D2 reaction. The data in Fig.~\ref{fig:ARPDOS} clearly indicate substantial changes of the valence band nature during the U2-D2 reaction: the top of the \emph{v}-band in D2 configuration is determined mostly by the contributions of \emph{up} and \emph{down} Si atoms (see atoms \emph{4} and \emph{5} in Fig.~\ref{fig:ARPDOS} \protect\subref{subfig:s-totd2u2grph} and \protect\subref{subfig:p-totd2u2grph}), in contrast to a single Si atom (hydrogenated Si atom in D2 configuration, Atom 4) contribution in U2. A similar reconstruction happens to the bottom of the \emph{c}-band (see  \emph{c}-band for atoms \emph{3}, \emph{4}, and \emph{5} in Fig.~\ref{fig:ARPDOS}). The transformation of the Si-O bond from dative to covalent pushes the oxygen related states out of the gap area (atom \emph{2} in Fig.~\ref{fig:ARPDOS} \protect\subref{subfig:p-totd2u2grph}). 

For the purposes of photosensitivity, it is instructive to consider further desorptive reactions of the ethanol on Si-surface. Reaction of U2-D4 results in more energetically favorable configuration. However the activation energy  for the U2-D4 reaction (TS2=1.1 eV) is much bigger then that for the U2-D2 reaction (see Fig.~\ref{fig:tssearch-grph}). The predicted configuration of D8 which is characterized by almost complete dissociation of the ethanol molecule through braking very strong C-C bonds has the lowest free energy (see Table~\ref{tab:energies-structures-tbl}). The D8 configuration is characterized by the ``bridge'' structure, achieved by breaking the C-C and O-H bonds of Ethanol during dissociative chemical reaction on Si(111) surface. The resulting atomic structure bridges two neighboring surface silicon atoms leading to the most stable configuration predicted in this work. However, this reaction has very high activation barrier (3.36 eV). This energy barrier prevents the ``bridge'' structure from being realized under normal conditions. This study clearly indicates that the D8 configuration-related process could only be possible under strong (\emph{e.g.} under laser power) excitations of the surface. 

Thus the analysis of our results indicates that the dissociation of Ethanol on Si(111) surface, that is accompanied by the creation of new covalent bonds, is characterized by substantial modifications of the valence band top (see the PDOS spectra) . Since top valence band electron transitions dominate in the optical absorption threshold, one can expect substantial modification of the photosensitivity of the Si-surface due to the ethanol adsorption. The predicted energy characteristics of the Ethanol adsorption on Si(111) surface could be used for photoinduced reactions, in particularly of the systems based on patterned Si-surfaces (e.g. PorSi) \cite{kolasinski06}. 

\section{Conclusions}
The equilibrium geometries of Ethanol adsorbed on bare (initially unreconstructed) Si~(111) surface are predicted by \emph{ab initio} pseudopotential method. Desorption of ethanol hydrogen and its reaction with the Si-dangling bond is shown to be a most probable product of the Ethanol reaction on Si(111) surface, consistent with the results previously obtained for Ethanol adsorption on Si~(100) surface. More energetically favorable configurations could only be achieved by overcoming substantial activation barriers. One of the most stable equilibrium structures found is the so called ``bridge'' structure, achieved by breaking two bonds, C-C and O-H of Ethanol during dissociative chemical reaction on Si(111) surface. The resulting atomic structure bridges two neighboring silicon surface atoms leading to the most stable configuration predicted in this work, however it is unlikely that this structure can be observed without substantial free energy being available in the system causing the D2 to become a trap state. The calculated energy barrier for the O-H bond clevage is significantly lower than that obtained for the Si~(100) surface. The absence of the Si-Si dimer, which is buckled during the undissociated adsorption, benefits the reaction by lowering the energy of the transition state. 

\section{Acknowledment}
This work is supported by STC MDITR National Science Foundation No. DMR-0120967 and NSF CREST-supplement grant No. HRD-0520208.

\end{document}